# Construction of Effective Interpolating Equation of State for One-and Two-Component Classical Plasma

V.K. Gryaznov  and  I.L. Iosilevskiy

*Moscow Institute of Physics and Technology*

In equilibrium plasma $s$-particle correlation functions $F_s(x_1,...x_s)$ must obey the normalizing conditions, which describe the local screening of an arbitrary charge.

$$\sum_{\alpha_{s+1}} n_{\alpha_{s+1}} e_{\alpha_{s+1}} \int \left[ \frac{F_{\alpha_1,...\alpha_{s+1}}}{F_{\alpha_1,...\alpha_s}} - 1 \right] d\mathbf{x}_{\alpha_{s+1}} = -\left( e_{\alpha_1} + ... + e_{\alpha_s} \right) \qquad (s = 0,1,2,...) \qquad (1)$$

Here summing is over all species, $n_i$ and $e_i$ – concentration and charge of specie $i$. These relations are followed from the theorems of existence for thermodynamic limit in grand canonical ensemble [1], and in particular from the fact that at this limit plasma tends to total electroneutrality ($\Sigma n_i e_i = 0$) independently on values of the chemical potentials $\{\mu_i\}$. It means that in thermodynamic limit

$$\det \left\| \frac{\partial n_i}{\partial \mu_j} \right\| = 0, \qquad \frac{\partial^k \sum n_i e_i}{\partial \mu_{\alpha_1}...\partial \mu_{\alpha_k}} = 0 \qquad (2)$$

The latter relation from (2) is equivalent to (1)*[)].

It is reasonable to suggest that condition of dipole and quadrupole moments screening, which is more stronger than (l), is valid in general case. I.e. it is necessary that there will be no diverging components ~ $1/r$, $1/r^2$, $1/r^3$ in average potential at large distances.

Independence of conditions (1,2) on degree of non-ideality allows one to correct effectively approximate $s$-particle correlation functions $F_s(x_1,...x_s)$, which are derived from the model constructions or from some variants of expansion over small parameter. Two following models can illustrate this statement.

## I. Classical one-component plasma model (OCP) with compensating background.

It is necessary that the binary correlation function $F_2(x)$ to obey two conditions:
I) Positiveness and exponential type of correlation functions

$$F_2(r) \geq 0; \qquad F_2(r) \sim \exp\{-\beta \Phi(r)\} \qquad (r \to 0) \qquad (3)$$

2) Screening (local electoneutrality condition [2,3])

$$\int [F_2(x) - 1] x^2 dx = -\Gamma_D \equiv -\frac{e^2}{kTr_D} \qquad (x \equiv r/r_D) \qquad (4)$$

Number of simple approximations uses binary correlation function $F_2(r)$ in linearized form:

$$F(x) = 1 - \exp(-Ax)/Bx \qquad \text{(LDH)}$$

here $A$ and $B$ depend on non-ideality parameter $\Gamma_D$. Condition (3) is violated in this linearized form (LDH). Well-known nonlinear form for $F_2(r)$ (DH) obey (3) but violates (4).

$$F(x) = \exp\{-\exp(-Ax)/Bx\} \qquad \text{(DH)}$$

---
*) Condition (1) is not satisfied in so-called *superposition approximation*: $F_{123} = F_{12}F_{23}F_{13}$.

The simplest correction, which combines simultaneous carrying out of (3) and (4) is following: only positive part of $F_2(x)$ is used, while $B$ becomes arbitrary norming quantity, which is derived from condition (4). For example the linearized approximation of Debye-Hückel (LDH) with $A = 1$ and $B = 1/\Gamma_D$ takes the following form (LDH-1)

$$F_2(x) = \begin{cases} 1 - (R/x)\exp(R - x) & x \geq R \equiv \sqrt[3]{1 - 3\Gamma_D} - 1, \\ 0 & x \leq R \equiv \sqrt[3]{1 - 3\Gamma_D} - 1 \end{cases} \quad (5)$$

$$\frac{U}{NkT} = \frac{1}{4}\left[(1 + 3\Gamma_D)^{2/3} - 1\right]$$

Four following approximations are exposed at Fig. 1 and 2 *before* and *after* such correction:
- Linearized and non-linearized Debye-Hückel approximations (notations LDH and DH)
- The approximation, which takes into account asymptotically dependence of amplitude $B^{-1}$ screening radius $A^{-1}$ in (LDH) from $\Gamma_D$ (notation MN) [4],
- The approximation equivalent to LDH, but in grand canonical ensemble (GDH) [5].

Results of the Monte-Carlo numerical simulations [6] were used as a standard in this comparison.

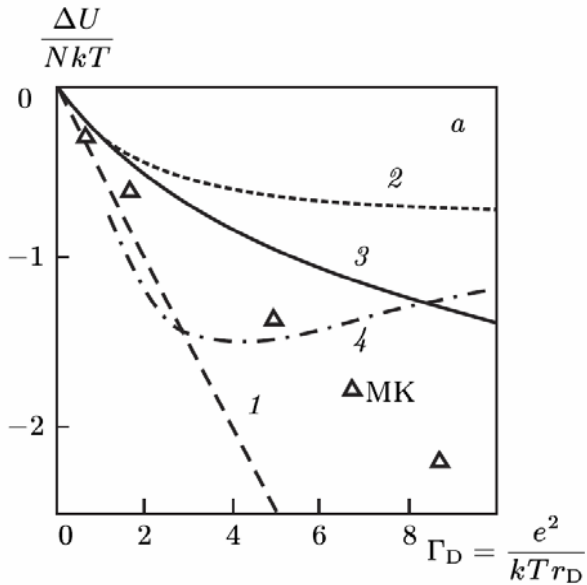
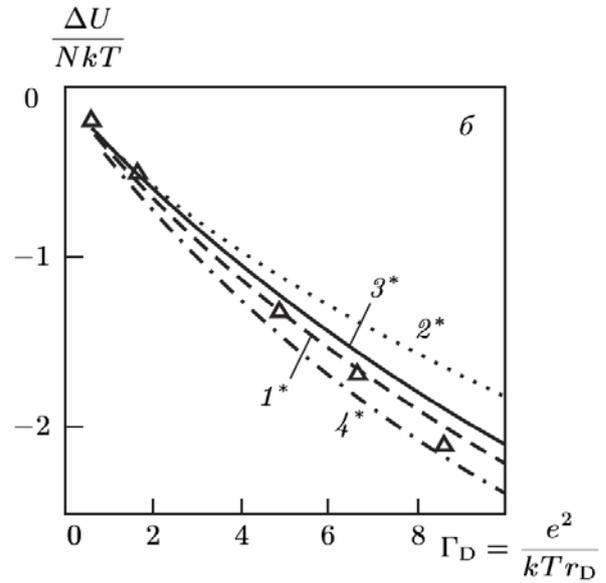

Fig. 1                                    Fig. 2

*1* – LDH, *2* – GDH, *3* – DH, *4* – MN [4], *MK* – Monte-Carlo [6]

Provided correction does not violate asymptotic behavior of corrected approximations at $\Gamma_D \ll 1$. At the same time this correction increases significantly agreement between corrected approximations and MC-etalon even when the true correlation functions became oscillating (wavy line on Fig. 2 and 3), while there are no such oscillations in normalized approximations like (5). Thus it may be concluded that the fact of such correction is more important then particular way of such correction.

From physical arguments [2] a constraint more stronger than (4), so-called "second momentum" condition, might be imposed on $F_2(x)$:

$$\int [F_2(x) - 1]x^4 dx = -6\Gamma_D \qquad (x \equiv r/r_D) \qquad (6)$$

It is followed from (4) and (6) that the oscillating behavior in $F(x)$ begin no later then at $\Gamma_D^* = (10)^{3/2}/3 \cong 10.5$. Moreover, it is possible to construct a simple oscillating approximation for $F(x)$ with two normalizing parameters to obey conditions (3, 4, 6) (notation – LDH-2)

$$F_2(x) = \begin{cases} 1 - (A/x)\exp(-x)\cos(Bx) & x \geq R, \\ 0 & x \leq R \end{cases} \quad (7)$$

The results are shown in Figure 3

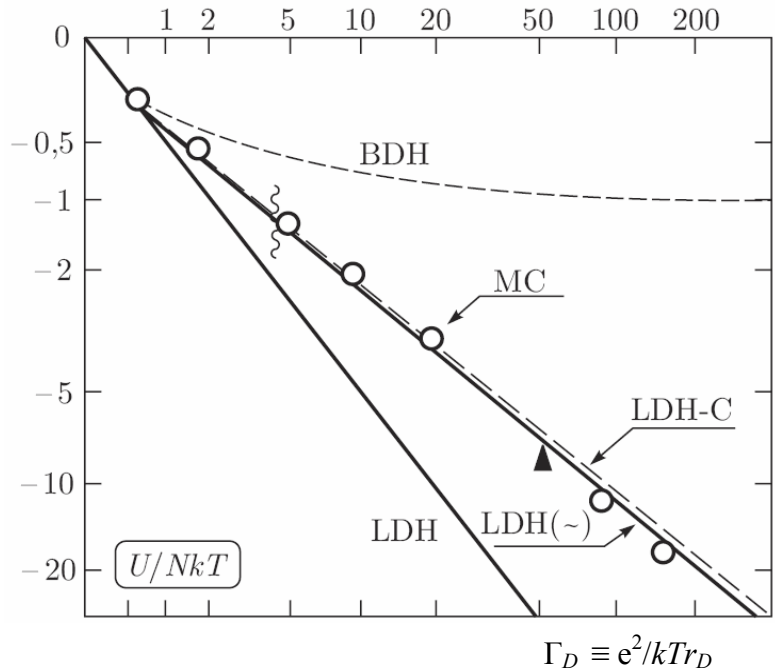

The condition (4) can be used as effective tool for checking the accuracy of the numerical solution of the integral equations for $F_2(x)$. For example, evident violation of this condition in solution of the equation *CHNC* in [7]*⁾ (see the table below) reveals incorrectness of the derived solution. One can conclude that there exists strong correlation between errors in normalization condition (4) and corresponding errors in resulting equation of state.

Table 1

| $(e^2/kTa)$   $a \equiv (4\pi n/3)^{-1/3}$ | 1/20 | 1/10 | 1/5 | 1/3 | 1 |
|---|---|---|---|---|---|
| $(1/\Gamma_D)\int_0^R [F_{CHNC}(x)-1]x^2 dx$ | 0.20 | 0.28 | 0.39 | 0.45 | 0.42 |
| $(1/\Gamma_D)\int_0^R [F_{CHNC}(x)-1]x^2 dx + (1/\Gamma_D)[F_{CHNC}(R)-1]\int_R^\infty R\cdot\exp(R-x)x dx$ | 0.82 | 0.64 | 0.57 | 0.57 | 0.44 |

-------------------------------

*⁾ In *CHNC* approximation conditions (3) and (4) are fulfilled automatically.

## 2. Classical two-component plasma model.

Let us consider the system of $2N$ charged particles with finite interaction potential at zero

$$\Phi_\pm(r) = \pm \frac{A}{r}[1-\exp(-\alpha r)] \qquad \Phi_\pm(0) \equiv \pm\Phi_0 = \pm A\alpha \tag{8}$$

From general physical arguments the binary correlation functions $F_\pm(r)$ are taken in the three-parametric form with three constants – $\psi_0$, $\lambda$ and $\omega$.

$$F_\pm(r) \equiv \exp\{\pm\psi(r)\} = \exp\{\pm\psi_0 e^{-\lambda r}[sh(\omega r)/\omega r]\} \tag{9}$$

The parameters $\psi_0$, $\lambda$ and $\omega$ ($\omega$ can be imaginary) can be chosen using following conditions:

1) The normalization condition of screening (local electroneutrality condition)

$$4\pi n \int (F_+ - F_-) r^2 dr = -1 \tag{10}$$

2) $$\psi_0 \equiv -\beta(\Phi_0 - \Delta I) = -\beta\left\{\Phi_0 - 4\pi n \int \Phi(r)[F_+ - F_-] r^2 dr\right\} \tag{11}$$

3) $$\left.\frac{d\psi_\pm(r)}{dr}\right|_{r=0} = \left.\frac{d}{dr}[-\beta\Phi_\pm(r)]\right|_{r=0} \tag{12}$$

At the weak non-ideality conditions $\{\Phi(r_{cp})/kT \ll 1\}$ the approximation (9-12) turns into so-called Glauberman-Yuhnovsky approximation (GU) [8]. Fig. 4 shows linearized (LGU) and non-linearized (GU) approximations (corresponding to [8]) in comparison with the presently constructed interpolating approximation (9-12).

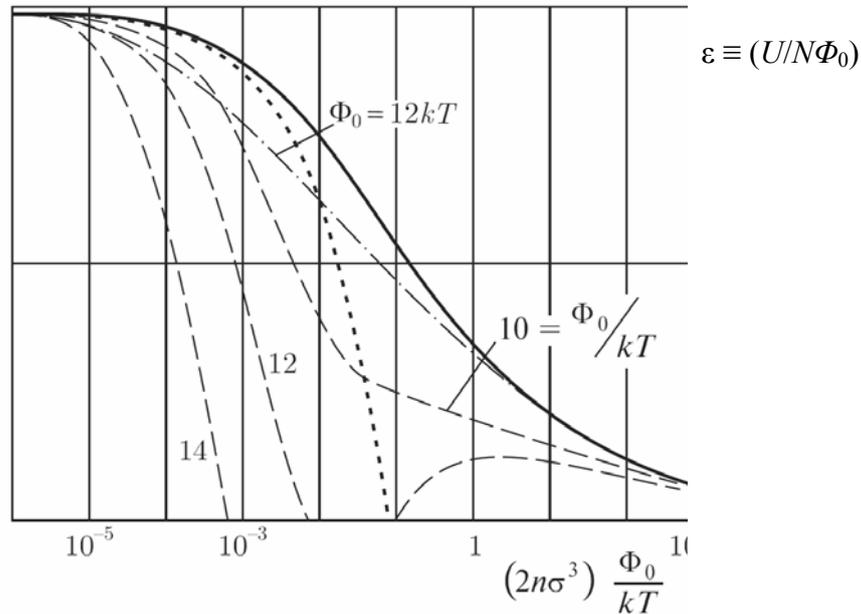

Figure 4

At sufficiently small densities linearized approximation LGU [8] is correct. As density increases the exponential dependence of the correlation function on short-range attraction at close distances (appearance of quasi-neutral pairs) should be taken into account. This effect is described by non-linear so-called "ladder" approximation (GU). However in this approximation (GU) the condition of local electro-neutrality (10) is violated. The screening charge is overestimated, which have consequences as the density increases. In this limit the value $\Delta I$ in (11) increases, so that the approximation (9-12) tends to the linearized approximation LGU.

It should be stressed that 1$^{st}$-order phase transition of gas-liquid type is presented in all three approximations. Their critical temperatures $T_c$ are listed in the table below:

Table 2

| Approximation | LGY | GY | (9 – 12) |
|---|---|---|---|
| $\Phi_0/kT_c$ | ~ 20 | ~ 12 | ~34 |